\documentclass{elsart1p}

\usepackage{amssymb,float}

\begin{document}

\begin{frontmatter}



\title{Shell-model studies on exotic nuclei around $^{132}$Sn}


\author[infn]{L. Coraggio}
\author[infn,dip]{A. Covello}
\author[infn]{A. Gargano}
\author[infn,dip]{N. Itaco}
\author[suny]{T. T. S. Kuo}
\address[infn]{Istituto Nazionale di Fisica Nucleare, Sezione di
  Napoli, \\
Complesso Universitario di Monte S. Angelo, I-80126 Napoli, Italy}
\address[dip]{Dipartimento di Scienze Fisiche, Universit\`a di Napoli
  ``Federico II'', \\ 
Complesso Universitario di Monte S. Angelo, I-80126 Napoli, Italy}
\address[suny]{Department of Physics and Astronomy, Stony Brook
  University, Stony Brook, NY 11794, USA}

\begin{abstract}
The study of exotic nuclei around $^{132}$Sn is a subject of current
experimental and theoretical interest. 
Experimental information for nuclei in the vicinity of $^{132}$Sn,
which have been long inaccessible to spectroscopic studies, is now
available thanks to new advanced facilities and techniques. 
The experimental data which have been now become available for these
neutron-rich nuclei may suggest a modification in the shell
structure.
They are, in fact, somewhat different from what one might expect by 
extrapolating the existing results for $N<82$, and as a possible
explanation a change in the single-proton level scheme has been
suggested. 
The latter would be caused by a more diffuse nuclear surface, and
could be seen as a precursor of major effects which should show up at
larger neutron excess. 
New data offer therefore the opportunity to test the shell model and
look for a possible evolution of shell structure when going toward
neutron drip line. 
This is stimulating shell-model studies in this region.
Here, we present an overview of recent shell-model studies 
of $^{132}$Sn neighbors, focusing attention on those calculations
employing realistic effective interactions. 
\end{abstract}

\end{frontmatter}

\section{Introduction}
\label{intro}
The region of nuclei around doubly-magic $^{132}$Sn is currently a
subject of a certain theoretical and experimental interest. 
New advanced facilities and techniques, such as the advent of
radioactive ion beams, allow to access to new data
\cite{Radford02,Shergur05a,Shergur05b}, that give the opportunity to
test theoretical models. 

In particular, it is important to study in this region, by way of
microscopic approaches, the possible evolution of shell structure when
going toward proton or neutron drip lines \cite{Otsuka04,Otsuka05}.
In fact, recent works evidence that in $N=82$ isotones the so-called
``shell-quenching'' phenomenon seems to play a fundamental role to 
reproduce the solar system $r-$process abudances ($N_{r,\odot}$)
\cite{Dillmann03}. 
This is stimulating shell-model studies in this region, focusing
particular attention on the two-body matrix elements (TBME) of the
residual interaction.

In recent years, the derivation of the shell-model TBME from realistic
nucleon-nucleon ($NN$) potentials has proved to be a reliable approach
to microscopic shell-model calculations \cite{Kartamyshev06,Covello07}.
The success achieved by these calculations in different mass regions
gives a clearcut answer to the long-standing problem of how accurate
description of nuclear structure properties can be provided by
realistic shell-model interactions, and opens the way to a more
fundamental approach to the nuclear shell model than the traditional
one, which makes use of empirical TBME with several parameters.

The paper is organized as follows. 
In Sec. II we give a short description of how the short-range
repulsion of a realistic $NN$ potential is renormalized before to be
employed in the derivation of the effective shell-model interaction.
In particular, we will focus attention on a recent approach, the
so-called $V_{\rm low-k}$ one \cite{Bogner02}, that allows to derive a 
low-momentum $NN$ realistic potential which preserves exactly the
onshell physics of the original potential.
In Sec. III a summary of the derivation of the shell-model effective
hamiltonian $H_{\rm eff}$ is presented with some details of our
calculations. 
In Sec. IV we present and discuss results obtained employing realistic
shell-model interactions for nuclei with valence nucleons outside the
doubly-magic $^{132}$Sn core. 
Some concluding remarks are given in Sec. V.

\section{The renormalization of the short-range repulsion}
\label{renorm}
Because of the strong repulsive core in the short-range region, which
is a feature common to all modern $NN$ potentials, the latter cannot
be used directly in the derivation of shell-model effective
interaction within a perturbative approach, that is the standard
procedure. 
So, as mentioned before, realistic potentials have to be renormalized
first. 
The standard way to renormalize the short-range repulsion is to resort
to the theory of the Brueckner reaction matrix $G$, that provides to
sum up the infinite series of ladder diagrams whose interaction
vertices are the $NN$ interaction itself.
The $G$ matrix is defined \cite{Krenciglowa76} by the integral equation:

\begin{equation}
G(\omega) = V_{NN} + V_{NN} Q_{2p} \frac{1}{\omega-Q_{2p} T Q_{2p}} Q_{2p}
G(\omega)~~,
\label{gmat}
\end{equation}

\noindent
where $V_{NN}$ represents the $NN$ potential, $T$ denotes the
two-nucleon kinetic energy, and $\omega$ is an energy variable (the
so-called starting energy), given by the energy of the in-coming nucleons. 
The two-body Pauli exclusion operator $Q_{2p}$ prevents double
counting, namely the intermediate states allowed for $G$ must be
outside of the chosen model space.
Thus the Pauli operator $Q_{2p}$ is dependent on the model space, and
so is the  $G$ matrix. 

Inspired by the effective theory, an alternative approach based on the
renormalization group (RG) has been recently introduced to renormalize
the short-range repulsion introducing a cutoff momentum $\Lambda$,
that decouples the fast and slow modes of the original $V_{NN}$. 

Let us now outline briefly the derivation of this low-momentum
potential $V_{\rm low-k}$ \cite{Bogner02}.
The repulsive core contained in $V_{NN}$ is smoothed by integrating
out the high-momentum modes of $V_{NN}$ down to $\Lambda$. 
This integration is carried out with the requirement that the deuteron
binding energy and low-energy phase shifts of $V_{NN}$ are preserved
by $V_{\rm low-k}$. 
This is achieved  by the following $T$-matrix
equivalence approach. 
We start from the half-on-shell $T$ matrix for $V_{NN}$ 
\begin{equation}
T(k',k,k^2) = V_{NN}(k',k) + \mathcal{P} \int _0 ^{\infty} q^2 dq
V_{NN}(k',q) \frac{1}{k^2-q^2} T(q,k,k^2 ) ~~,
\end{equation}

\noindent
where $\mathcal{P}$ denotes the principal value and  $k,~k'$, and $q$
stand for the relative momenta. 
The effective low-momentum $T$ matrix is then defined by
\begin{equation}
T_{\rm low-k }(p',p,p^2) = V_{\rm low-k }(p',p) + \mathcal{P} \int _0
^{\Lambda} q^2 dq  V_{\rm low-k }(p',q) \frac{1}{p^2-q^2} T_{\rm 
low-k} (q,p,p^2) ~~,
\end{equation}

\noindent
where the intermediate state momentum $q$ is integrated from 0 to the
momentum space cutoff $\Lambda$ and $(p',p) \leq \Lambda$. 
The above $T$ matrices are required to satisfy the condition 
\begin{equation}
T(p',p,p^2)= T_{\rm low-k }(p',p,p^2) \, ; ~~ (p',p) \leq \Lambda \,.
\end{equation}

The above equations define the effective low-momentum interaction 
$V_{\rm low-k}$, and it has been shown \cite{Bogner02} that they are
satisfied by the solution:
\begin{equation}
V_{\rm low-k} = \hat{Q} - \hat{Q}' \int \hat{Q} + \hat{Q}' \int
\hat{Q} \int \hat{Q} - \hat{Q}' \int \hat{Q} \int \hat{Q} \int \hat{Q}
+ ~...~~,
\label{vlowk}
\end{equation}

\noindent
which is the well known Kuo-Lee-Ratcliff (KLR) folded-diagram
expansion \cite{Kuo71,Kuo90}, originally designed for constructing
shell-model effective interactions.
In Eq. (\ref{vlowk}) $\hat{Q}$ is an irreducible vertex function
whose intermediate states are all beyond $\Lambda$ and $\hat{Q}'$ is
obtained by removing from $\hat{Q}$ its terms first order in the
interaction $V_{NN}$. 
In addition to the preservation of the half-on-shell $T$ matrix, which
implies preservation of the phase shifts, this $V_{\rm low-k}$
preserves the deuteron binding energy, since eigenvalues are preserved
by the KLR effective interaction. 
For any value of $\Lambda$, the low-momentum potential of
Eq. (\ref{vlowk}) can be calculated very accurately using iteration
methods. 
Our calculation of $V_{\rm low-k}$ is performed by employing the
iteration method proposed in \cite{Andreozzi96}, which is based on the 
Lee-Suzuki similarity transformation \cite{Suzuki80}. 

The main result is that $V_{\rm low-k}$ is a smooth potential which
preserves exactly the onshell properties of the original $V_{NN}$, and
is suitable to be used directly in nuclear structure calculations.
In the past few years, $V_{\rm low-k}$ has been fruitfully employed
in microscopic calculations within different perturbative frameworks
such as the realistic shell model 
\cite{Coraggio02,Coraggio04,Coraggio05a,Coraggio06a}, 
the Goldstone expansion for doubly closed-shell nuclei 
\cite{Coraggio03,Coraggio05b,Coraggio06b}, and the Hartree-Fock theory
for nuclear matter calculations \cite{Sedrakian03,Bogner05}.

\section{The derivation of the shell-model effective potential}
\label{effint}
In the framework of the shell model, an auxiliary one-body potential
$U$ is introduced in order to break up the nuclear hamiltonian as the
sum of a one-body component $H_0$, which describes the independent
motion of the nucleons, and a residual interaction $H_1$:

\begin{equation}
H=\sum_{i=1}^{A} \frac{p_i^2}{2m} + \sum_{i<j} V_{ij} = T + V =
(T+U)+(V-U)= H_{0}+H_{1}~~.
\label{smham}
\end{equation}

Once $H_0$ has been introduced, a reduced model space is defined in
terms of a finite subset of $H_0$'s eigenvectors. 
The diagonalization of the many-body hamiltonian (\ref{smham}) in an
infinite Hilbert space, that is obviously unfeasible, is then
reduced to the solution of an eigenvalue problem for an effective 
hamiltonian $H_{\rm eff}$ in a finite space.

The standard approach is to derive $H_{\rm eff}$ by way of the
time-dependent perturbation theory \cite{Kuo71,Kuo90}.
Namely, $H_{\rm eff}$ is expressed through the KLR folded-diagram
expansion in terms of the vertex function $\hat{Q}$-box, which is
composed of irreducible valence-linked diagrams. 
The $\hat{Q}$-box is composed of one- and two-body Goldstone diagrams
through a certain order in $V$ \cite{Kuo81}, where $V$ is the
renormalized input potential.
Once the $\hat{Q}$-box has been calculated, the series of the folded
diagrams is summed up to all orders using the Lee-Suzuki iteration
method \cite{Suzuki80}.

The hamiltonian $H_{\rm eff}$ contains one-body contributions, which
represent the effective single-particle (SP) energies.
In realistic shell-model calculations it is customary to use a
subtraction procedure \cite{Shurpin83} so that only the two-body terms
of $H_{\rm eff}$ are retained - the effective interaction $V_{\rm
eff}$ - and the SP energies are taken from the experimental data.
This is what has also been done in the calculations reported in the
following section.
The single-proton and single-neutron energies have been taken from the
experimental spectra of $^{133}$Sb and $^{133}$Sn \cite{nndc}, that
can be described just as one proton and one neutron diving in the mean
field generated by the $^{132}$Sn doubly-closed core.
For sake of completeness, it is important to mention that
experimentally the proton $2s_{1/2}$ level is missing and the neutron
$0i_{13/2}$ level is unbound.
The analysis of the first excited $J^{\pi}=10^+$ in $^{134}$Sb allows
to estimate the SP energy of neutron $0i_{13/2}$ level to be $2.694
\pm 0.2$ MeV \cite{Urban99}, while a study of odd $N=82$ isotones
suggests that the SP proton $2s_{1/2}$ level should lie around $2.8$ MeV
excitation energy \cite{Andreozzi97}.

\section{Realistic shell-model calculations}
\label{results}
Here, we present some selected results employing realistic shell-model
$V_{\rm eff}$s for nuclei with valence nucleons outside $^{132}$Sn
core. 
The $V_{\rm eff}$s are derived from the CD-Bonn realistic $NN$
potential \cite{Machleidt01} using as renormalization approach both
the $G$-matrix and the $V_{\rm low-k}$ ones. 
Calculations employing a $G$-matrix derived from the CD-Bonn potential
have been widely performed by the Oslo group and their coworkers 
\cite{Shergur05a,Shergur05b,Kartamyshev06,Shergur02,Brown05} with a
remarkable success. 
The $V_{\rm eff}$ is derived within the folded-diagram approach
described in section \ref{effint}, including in the $\hat{Q}$-box
diagrams up to the third order in $G$ and intermediate states with at
most $2 \hbar \omega$ excitation energy \cite{Hjorth96}. 

The region of nuclei around $^{132}$Sn core has beeen studied in recent
years also by the Naples-Stony Brook group within the framework of the
$V_{\rm low-k}$ renormalization procedure. 
In such a case, the shell-model effective interaction is obtained
starting from a $V_{\rm low-k}$ derived from the CD-Bonn potential,
with a cutoff momentum $\Lambda=2.2$ fm$^{-1}$, that is a value able to
preserve all the two-nucleon physics of the original potential up to
the anelastic threshold and small enough so to give a reasonably smooth
potential. 
Then the $\hat{Q}$-box has been calculated including diagrams up to
second order in $V_{\rm low-k}$ using intermediate states composed of
all possible hole states and particle states restricted to the five
shells above the $^{132}$Sn Fermi surface.
This guarantees the stability of the $V_{\rm eff}$ TBME when
increasing the number of intermediate particle states.

In both approaches an harmonic-oscillator basis with an oscillator
parameter $\hbar \omega=7.88$ MeV has been employed.

Let us now come to the results of the calculations and their
comparison with experimental data.

First, it is worth to point out that a fundamental test for the
reliability of the matrix elements of $V_{\rm eff}$ are the systems
with two valence nucleons outside di closed-shell core. 
In the present case, the test is the reproduction of the spectra of
$^{134}$Te - two protons outside $^{132}$Sn - $^{134}$Sn, and
$^{134}$Sb, that are two neutrons and one proton and one neutron
outside $^{132}$Sn, respectively.

In Tables \ref{134te},\ref{134sn},\ref{134sb} the experimental and
theoretical low-lying spectra of $^{134}$Te, $^{134}$Sn, $^{134}$Sb
are reported.

\begin{table}[H]
\begin{center}
\caption{Experimental energy levels up to 3 MeV \cite{nndc} for
  $^{134}$Te compared to the calculation with CD-Bonn potential
  through $G$-matrix \cite{Brown05} and $V_{\rm low-k}$
  \cite{Covello07} renormalization procedures, respectively.} 
\begin{tabular}{ccccccc}
\hline
 $J^{\pi}$ & ~ & Experiment & ~ & $G$-matrix & ~ & $V_{\rm low-k}$ \\
\hline
$0^+_1$ & ~ &  0.0  & ~ &  0.0  & ~ &  0.0  \\
$2^+_1$ & ~ &  1.28 & ~ &  1.21 & ~ &  1.33 \\
$4^+_1$ & ~ &  1.57 & ~ &  1.48 & ~ &  1.61 \\
$6^+_1$ & ~ &  1.69 & ~ &  1.61 & ~ &  1.75 \\
$6^+_2$ & ~ &  2.40 & ~ &  2.17 & ~ &  2.45 \\
$2^+_2$ & ~ &  2.46 & ~ &  2.45 & ~ &  2.67 \\
$4^+_2$ & ~ &  2.55 & ~ &  2.45 & ~ &  2.63 \\
$1^+_1$ & ~ &  2.63 & ~ &  2.41 & ~ &  2.67 \\
$3^+_1$ & ~ &  2.68 & ~ &  2.54 & ~ &  2.68 \\
$5^+_1$ & ~ &  2.73 & ~ &  2.54 & ~ &  2.68 \\
$2^+_3$ & ~ &  2.93 & ~ &  3.06 & ~ &  3.27 \\
\hline
\end{tabular}
\label{134te}
\end{center}
\end{table}

\begin{table}[H]
\begin{center}
\caption{Experimental observed energy levels \cite{nndc} for
  $^{134}$Sn compared to the calculation with CD-Bonn potential
  through $G$-matrix \cite{Kartamyshev06} and $V_{\rm low-k}$
  \cite{Covello07} renormalization procedures, respectively.} 
\begin{tabular}{ccccccc}
\hline
 $J^{\pi}$ & ~ & Experiment & ~ & $G$-matrix & ~ & $V_{\rm low-k}$ \\
\hline
$0^+_1$ & ~ &  0.0   & ~ &  0.0   & ~ &  0.0   \\
$2^+_1$ & ~ &  0.726 & ~ &  0.775 & ~ &  0.733 \\
$4^+_1$ & ~ &  1.073 & ~ &  1.116 & ~ &  1.016 \\
$6^+_1$ & ~ &  1.247 & ~ &  1.258 & ~ &  1.125 \\
$8^+_1$ & ~ &  2.509 & ~ &  2.463 & ~ &  2.545 \\
\hline
\end{tabular}
\label{134sn}
\end{center}
\end{table}

From the inspection of Tables \ref{134te},\ref{134sn}, it is evident that
the agreement between theory and experiment is very good for the
identical particle channel, both with $G$-matrix and $V_{\rm low-k}$
approach.
It is worth to note that theoretical results do not differ so much
when using the two different renormalization procedures.

\begin{table}[ht]
\begin{center}
\caption{Experimental energy levels up to 1 MeV \cite{Shergur05a} for
  $^{134}$Sb compared to the calculation with CD-Bonn potential
  through $G$-matrix \cite{Shergur05a} and $V_{\rm low-k}$
  \cite{Coraggio06a} renormalization procedures, respectively.} 
\begin{tabular}{ccccccc}
\hline
 $J^{\pi}$ & ~ & Experiment & ~ & $G$-matrix & ~ & $V_{\rm low-k}$ \\
\hline
$0^-_1$ & ~ &  0.0   & ~ &  0.0   & ~ &  0.0   \\
$1^-_1$ & ~ &  0.013 & ~ &  0.329 & ~ &  0.052 \\
$7^-_1$ & ~ &  0.279 & ~ &  0.392 & ~ &  0.407 \\
$2^-_1$ & ~ &  0.330 & ~ &  0.406 & ~ &  0.385 \\
$3^-_2$ & ~ &  0.383 & ~ &  0.581 & ~ &  0.419 \\
$5^-_1$ & ~ &  0.442 & ~ &  0.604 & ~ &  0.494 \\
$4^-_1$ & ~ &  0.555 & ~ &  0.710 & ~ &  0.621 \\
$6^-_1$ & ~ &  0.617 & ~ &  0.849 & ~ &  0.727 \\
$1^-_2$ & ~ &  0.885 & ~ &  1.268 & ~ &  0.868 \\
$2^-_2$ & ~ &  0.935 & ~ &  1.051 & ~ &  0.958 \\
\hline
\end{tabular}
\label{134sb}
\end{center}
\end{table}

The study of $^{134}$Sb low-lying spectrum evidences a differ
situation respect the identical-particle case.
A less better agreement with experimental data is obtained employing
the $V_{\rm eff}$ TBME derived with the $G$-matrix approach.
This deficiency is also propagated in the theoretical spectrum of
$^{135}$Sb obtained with $G$-matrix renormalization procedure, as it
can be seen in Table \ref{135sb}.
In Ref.\cite{Shergur05b}, in order to obtain a better agreement with
experimental spectrum of $^{135}$Sb, it was pointed out that a
downshift of the proton $d_{5/2}$ level with respect to the $g_{7/2}$
one by 300 keV, as a possible collective influence of a neutron skin
\cite{Shergur02}, turned out to be necessary, but did not help for
$^{134}$Sb \cite{Shergur02,Shergur05a}.

We have verified that the differences between the results obtained
with $G$-matrix and those with $V_{\rm low-k}$, that are in quite good
agreement with experiment, should be traced mainly to the different
dimension of the intermediate state space.
In fact, it has been found that including intermediate states only up
to $2 \hbar \omega$ excitation energy in the calculation of the
second-order $\hat{Q}$-box, with the $V_{\rm low-k}$ as input
potential, the results become similar, as in the case of the
identical-particle channel.

\begin{table}[H]
\begin{center}
\caption{Experimental energy levels up to 1 MeV \cite{Shergur05b} for
  $^{135}$Sb compared to the calculation with CD-Bonn potential
  through $G$-matrix \cite{Shergur05b} and $V_{\rm low-k}$
  \cite{Coraggio05c} renormalization procedures, respectively.} 
\begin{tabular}{ccccccc}
\hline
 $J^{\pi}$ & ~ & Experiment & ~ & $G$-matrix & ~ & $V_{\rm low-k}$ \\
\hline
$7/2^+_1$  & ~ &  0.0   & ~ &  0.0   & ~ &  0.0   \\
$5/2^+_1$  & ~ &  0.282 & ~ &  0.527 & ~ &  0.391 \\
$3/2^+_1$  & ~ &  0.440 & ~ &  0.438 & ~ &  0.509 \\
$11/2^+_1$ & ~ &  0.707 & ~ &  0.662 & ~ &  0.750 \\
$9/2^+_1$  & ~ &  0.798 & ~ &  0.947 & ~ &  0.813 \\
$7/2^+_2$  & ~ &  1.014 & ~ &  1.135 & ~ &  0.938 \\
$9/2^+_2$  & ~ &  1.027 & ~ &  1.165 & ~ &  1.108 \\
\hline
\end{tabular}
\label{135sb}
\end{center}
\end{table}

This confirms the fact that results do not depend so much on the
renormalization technique.
However, it is worth to note that, because $V_{\rm low-k}$ does not
depend on the Pauli-blocking operator $Q$ as the $G$-matrix, using
$V_{\rm low-k}$ one can easily employ a larger number of intermediate
states and obtaining consequently better results.

\section{Concluding remarks}
Here, we have presented selected results of some shell-model studies
of nuclei with valence-nucleons outside doubly-closed shell core
$^{132}$Sn, where realistic effective shell-model interactions have
been employed.
In particular, we have focused the attention on few valence-nucleons
nuclei which are most appropriate for a stringent test of the two-body
matrix elements.
The latter have been derived by means of a $\hat{Q}$-box
folded-diagram method from the CD-Bonn potential, renormalized both by
use of the $G$-matrix and $V_{\rm low-k}$ approaches.
Results are in a good agreement with experiment and, in particular, do
not depend strongly on the renormalization technique employed, except
a slightly difference in the neutron-proton interaction.
To conclude, ...




\begin{thebibliography}{99}
\bibitem{Radford02} D. C. Radford {\em et al.}, Phys. Rev. Lett. {\bf
  88}, 222501 (2002).
\bibitem{Shergur05a} J. Shergur {\em et al.}, Phys. Rev.  {\bf
  71}, 064321 (2005).
\bibitem{Shergur05b} J. Shergur {\em et al.}, Phys. Rev.  {\bf
  72}, 024305 (2005).
\bibitem{Otsuka04} T. Otsuka, Nucl. Phys. A {\bf 734}, 365 (2004).
\bibitem{Otsuka05} T. Otsuka, T. Suzuki, R. Fujimoto, H. Grawe, and
  Y. Akaishi, Phys. Rev. Lett. {\bf 95}, 232502 (2005).
\bibitem{Dillmann03} I. Dillmann {\em et al.}, Phys. Rev. Lett. {\bf
  91}, 162503 (2003).
\bibitem{Kartamyshev06} M. P. Kartamyshev, T. Engeland,
  M. Hjorth-Jensen, and E. Osnes, arXiv:nucl-th/0610017 (2006) and
  references therein.
\bibitem{Covello07} A. Covello, L. Coraggio, A. Gargano, and N. Itaco,
  Prog. Part. Nucl. Phys. {\bf 59}, 401 (2007) and references therein.
\bibitem{Bogner02}  S. Bogner, T. T. S. Kuo, L. Coraggio, A. Covello,
  and N. Itaco, Phys. Rev. C {\bf 65}, 051301(R) (2002).
\bibitem{Krenciglowa76}  E. M. Krenciglowa, C. L. Kung, and
  T. T. S. Kuo, Ann. Phys. (NY) {\bf 101}, 154 (1976).
\bibitem{Kuo71} T. T. S. Kuo, S. Y. Lee, and K. F. Ratcliff,
  Nucl. Phys A {\bf 176}, 65 (1971).
\bibitem{Kuo90} T. T. S. Kuo and E. Osnes {\it Lecture Notes in
  Physics} vol. {\bf 364} (Springer-Verlag, Berlin) (1990).
\bibitem{Andreozzi96} F. Andreozzi Phys. Rev. C {\bf 54}, 684 (1996).
\bibitem{Suzuki80} K. Suzuki and S. Y. Lee, Prog. Theor. Phys. {\bf
  64}, 2091 (1980).
\bibitem{Coraggio02}  L. Coraggio, A. Covello, A. Gargano, and
  N. Itaco, Phys. Rev. C {\bf 66}, 064311 (2002).
\bibitem{Coraggio04}  L. Coraggio, A. Covello, A. Gargano, and
  N. Itaco, Phys. Rev. C {\bf 70}, 034310 (2004).
\bibitem{Coraggio05a}  L. Coraggio and N. Itaco, Phys. Lett. B {\bf
  616}, 43 (2005).
\bibitem{Coraggio06a}  L. Coraggio, A. Covello, A. Gargano, and
  N. Itaco, Phys. Rev. C {\bf 73}, 031302(R) (2006).
\bibitem{Coraggio03}  L. Coraggio, N. Itaco, A. Covello, A. Gargano,
  and T. T. S. Kuo, Phys. Rev. C {\bf 68}, 034320 (2003).
\bibitem{Coraggio05b}  L. Coraggio, A. Covello, A. Gargano, N. Itaco,
  T. T. S. Kuo, and R. Machleidt, Phys. Rev. C {\bf 71}, 014307 (2005).
\bibitem{Coraggio06b}  L. Coraggio, A. Covello, A. Gargano, N. Itaco,
  and T. T. S. Kuo, Phys. Rev. C {\bf 73}, 014304 (2006).
\bibitem{Sedrakian03}  A. Sedrakian, T. T. S. Kuo, H. M{\"u}ther, and
  P. Schuck, Phys. Lett. B {\bf 576}, 68 (2003).
\bibitem{Bogner05} S. K. Bogner, A. Schwenk, R. J. Furnstahl, and
  A. Nogga, Nucl. Phys. A {\bf 763}, 59 (2005).
\bibitem{Kuo81} T. T. S. Kuo, J. Shurpin, K. C. Tam, E. Osnes, and
  P. J. Ellis, Ann. Phys. (NY) {\bf 132}, 237 (1981).
\bibitem{Shurpin83} J. Shurpin, T. T. S. Kuo, and D. Strottman,
  Nucl. Phys A {\bf 408}, 310 (1983).
\bibitem{nndc} Data extracted using the NNDC On-line Data Service from
  the ENSDF database, version of July 20, 2007.
\bibitem{Urban99} W. Urban {\em et al.}, Eur. Phys. J. A {\bf 5}, 239
  (1999).
\bibitem{Andreozzi97} F. Andreozzi. L. Coraggio, A. Covello,
  A. Gargano, T. T. S. Kuo, and A. Porrino, Phys. Rev. C {\bf56}, R16
  (1997).
\bibitem{Machleidt01} R. Machleidt, Phys. Rev. C {\bf 63}, 024001 (2001).
\bibitem{Shergur02} J. Shergur {\em et al.}, Phys. Rev. C {\bf 65},
  034313 (2002)
\bibitem{Brown05} B. A. Brown, N. J. Stone, I. S. Towner, and
M. Hjorth-Jensen, Phys. Rev. C {\bf 71}, 044317 (2005).
\bibitem{Hjorth96} M. Hjorth-Jensen, H. M{\"u}ther, and A. Polls,
  J. Phys. G {\bf 22}, 321 (1996).
\bibitem{Coraggio05c}  L. Coraggio, A. Covello, A. Gargano, and
  N. Itaco, Phys. Rev. C {\bf 72}, 057302 (2005).





\end{thebibliography}
\end{document}